\documentclass[5p,times]{elsarticle}

\usepackage{amssymb}
\usepackage{graphicx}
\usepackage{subcaption}
\usepackage{array}
\usepackage{multirow}
\usepackage{algorithm}
\usepackage{algpseudocode}
\usepackage{amsmath}
\usepackage{booktabs}
\usepackage{tablefootnote}
\usepackage{listings}
\usepackage[switch]{lineno}

\journal{Knowledge-Based Systems}

\begin{document}

\begin{frontmatter}



\title{VulnSense: Efficient Vulnerability Detection in Ethereum Smart Contracts by Multimodal Learning with Graph Neural Network and Language Model}

\author[inst1,inst2]{Phan The Duy}\ead{duypt@uit.edu.vn}
\author[inst1,inst2]{Nghi Hoang Khoa}\ead{khoanh@uit.edu.vn}
\author[inst1,inst2]{Nguyen Huu Quyen}\ead{quyennh@uit.edu.vn}
\author[inst1,inst2]{Le Cong Trinh}\ead{19522404@gm.uit.edu.vn}
\author[inst1,inst2]{Vu Trung Kien}\ead{19521722@gm.uit.edu.vn}
\author[inst1,inst2]{Trinh Minh Hoang}\ead{19521548@gm.uit.edu.vn}
\author[inst1,inst2]{Van-Hau Pham}\ead{haupv@uit.edu.vn}

\affiliation[inst1]{organization={Information Security Laboratory, University of Information Technology},
            city={Ho Chi Minh city},
            country={Vietnam}}

\affiliation[inst2]{organization={Vietnam National University Ho Chi Minh City},
            city={Hochiminh City},
            country={Vietnam}}

\begin{abstract}
This paper presents VulnSense framework, a comprehensive approach to efficiently detect vulnerabilities in Ethereum smart contracts using a multimodal learning approach on graph-based and natural language processing (NLP) models. Our proposed framework combines three types of features from smart contracts comprising source code, opcode sequences, and control flow graph (CFG) extracted from bytecode. We employ Bidirectional Encoder Representations from Transformers (BERT), Bidirectional Long Short-Term Memory (BiLSTM) and Graph Neural Network (GNN) models to extract and analyze these features. The final layer of our multimodal approach consists of a fully connected layer used to predict vulnerabilities in Ethereum smart contracts. Addressing limitations of existing vulnerability detection methods relying on single-feature or single-model deep learning techniques, our method surpasses accuracy and effectiveness constraints. We assess VulnSense using a collection of 1.769 smart contracts derived from the combination of three datasets: Curated, SolidiFI-Benchmark, and Smartbugs Wild. We then make a comparison with various unimodal and multimodal learning techniques contributed by GNN, BiLSTM and BERT architectures. The experimental outcomes demonstrate the superior performance of our proposed approach, achieving an average accuracy of 77.96\% across all three categories of vulnerable smart contracts.
\end{abstract}




\end{frontmatter}


\section{Introduction}
The Blockchain keyword has become increasingly more popular in the era of Industry 4.0 with many applications for a variety of purposes, both good and bad. For instance, in the field of finance, Blockchain is utilized to create new, faster, and more secure payment systems, examples of which include Bitcoin and Ethereum. However, Blockchain can also be exploited for money laundering, as it enables anonymous money transfers, as exemplified by cases like Silk Road \cite{Ghimiray_2023}. The number of keywords associated with blockchain is growing rapidly, reflecting the increasing interest in this technology. A typical example is the smart contract deployed on Ethereum. Smart contracts are programmed in Solidity, which is a new language that has been developed in recent years. When deployed on a blockchain system, smart contracts often execute transactions related to cryptocurrency, specifically the ether (ETH) token. However, the smart contracts still have many vulnerabilities, which have been pointed out by Zou et al. \cite{wei_chall_opport}. Alongside the immutable and transparent properties of Blockchain, the presence of vulnerabilities in smart contracts deployed within the Blockchain ecosystem enables attackers to exploit flawed smart contracts, thereby affecting the assets of individuals, organizations, as well as the stability within the Blockchain ecosystem.

In more detail, the DAO attack \cite{Meh_dao_attack} presented by Mehar et al. is a clear example  of the severity of the vulnerabilities, as it resulted in significant losses up to \$50 million. To solve the issues, Kushwaha et al. \cite{Kush_ETH_review} conducted a research survey on the different types of vulnerabilities in smart contracts and provided an overview of the existing tools for detecting and analyzing these vulnerabilities. Developers have created a number of tools to detect vulnerabilities in smart contracts source code, such as Oyente \cite{bar_smart2contract}, Slither \cite{Kush_ETH_review}, Conkas \cite{Nveloso}, Mythril \cite{Consensys}, Securify \cite{Tsan_secfy}, etc. These tools use static and dynamic analysis for seeking vulnerabilities, but they may not cover all execution paths, leading to false negatives. Additionally, exploring all execution paths in complex smart contracts can be time-consuming. Current endeavors in contract security analysis heavily depend on predefined rules established by specialists, a process that demands significant labor and lacks scalability.
 
 Meanwhile, the emergence of Machine Learning (ML) methods in the detection of vulnerabilities in software has also been explored. This is also applicable to smart contracts, where numerous tools and research are developed to identify security bugs, such as ESCORT by Lutz \cite{lutz2021escort}, ContractWard by Wang \cite{wang_contractward} and Qian \cite{qian_auto_squen_mol}. The ML-based methods have significantly improved performance over static and dynamic analysis methods, as indicated in the study by Jiang \cite{jiang_sur_smartcontr}. 

 However, the current studies do exhibit certain limitations, primarily centered around the utilization of only a singular type of feature from the smart contract as the input for ML models. To elaborate, it is noteworthy that a smart contract's representation and subsequent analysis can be approached through its source code, employing techniques such as NLP, as demonstrated in the study conducted by Khodadadi et al. \cite{khodadadi2023hymo}. Conversely, an alternative approach, as showcased by Chen et al. \cite{Chen_defec_check_2022}, involves the usage of the runtime bytecode of a smart contract published on the Ethereum blockchain. Additionally, Wang and colleagues \cite{wang_contractward} addressed vulnerability detection using opcodes extracted through the employment of the Solc tool \cite{Solidity} (Solidity compiler), based on either the contract's source code or bytecode.

In practical terms, these methodologies fall under the categorization of unimodal or monomodal models, designed to exclusively handle one distinct type of data feature. Extensively investigated and proven beneficial in domains such as computer vision, natural language processing, and network security, these unimodal models do exhibit impressive performance characteristics. However, their inherent drawback lies in their limited perspective, resulting from their exclusive focus on singular data attributes, which lack the potential characteristics for more in-depth analysis.

This limitation has prompted the emergence of multimodal models, which offer a more comprehensive outlook on data objects. The works of Jabeen and colleagues \cite{summaira2022review}, Tadas et al. \cite{Tadas2019_multimodal_survey}, Nam et al. \cite{nam2023survey}, and Xu \cite{xu2023_multimodal_trans_sur} underscore this trend. Specifically, multimodal learning harnesses distinct ML models, each accommodating diverse input types extracted from an object. This approach facilitates the acquisition of holistic and intricate representations of the object, a concerted effort to surmount the limitations posed by unimodal models. By leveraging multiple input sources, multimodal models endeavor to enrich the understanding of the analyzed data objects, resulting in more comprehensive and accurate outcomes.

 Recently, multimodal vulnerability detection models for smart contracts have emerged as a new research area, combining different techniques to process diverse data, including source code, bytecode and opcodes, to enhance the accuracy and reliability of AI systems. Numerous studies have demonstrated the effectiveness of using multimodal deep learning models to detect vulnerabilities in smart contracts. For instance, Yang et al. \cite{novel_ai_smartcontract} proposed a multimodal AI model that combines source code, bytecode, and execution traces to detect vulnerabilities in smart contracts with high accuracy. Chen et al. \cite{khodadadi2023hymo} proposed a new hybrid multimodal model called HyMo Framework, which combines static and dynamic analysis techniques to detect vulnerabilities in smart contracts. Their framework uses multiple methods and outperforms other methods on several test datasets.
 
Recognizing that these features accurately reflect smart contracts and the potential of multimodal learning, we employ a multimodal approach to build a vulnerability detection tool for smart contracts called VulnSense. Different features can provide unique insights into vulnerabilities on smart contracts. Source code offers a high-level understanding of contract logic, bytecode reveals low-level execution details, and opcode sequences capture the execution flow. By fusing these features, the model can extract a richer set of features, potentially leading to more accurate detection of vulnerabilities. The main contributions of this paper are summarized as follows:

\begin{itemize}
    \item First, we propose a multimodal learning approach consisting of BERT, BiLSTM and GNN to analyze the smart contract under multi-view strategy by leveraging the capability of NLP algorithms, corresponding to three type of features, including source code, opcodes, and CFG generated from bytecode. 

    \item Then, we extract and leverage three types of features from smart contracts to make a comprehensive feature fusion. More specifics, our smart contract representations which are created from real-world smart contract datasets, including Smartbugs Curated, SolidiFI-Benchmark and Smartbugs Wild, can help model to capture semantic relationships of characteristics in the phase of analysis.
    
    \item Finally, we evaluate the performance of VulnSense framework on the real-world vulnerable smart contracts to indicate the capability of detecting security defects such as Reentrancy, Arithmetic on smart contracts. Additionally, we also compare our framework with a unimodal models and other multimodal ones to prove the superior effectiveness of VulnSense.
\end{itemize}
The remaining sections of this article are constructed as follows. Section \ref{related_work} introduces some related works in adversarial attacks and countermeasures. The section \ref{Background} gives a brief background of applied components. Next, the threat model and methodology are discussed in section \ref{methodology}. Section \ref{experiments} describes the experimental settings and scenarios with the result analysis of our work. Finally, we conclude the paper in section \ref{conclusion}.

\section{Background} \label{Background}
\subsection{Bytecode of Smart Contracts}
Bytecode is a sequence of hexadecimal machine instructions generated from high-level programming languages such as C/C++, Python, and similarly, Solidity. In the context of deploying smart contracts using Solidity, bytecode serves as the compiled version of the smart contract's source code and is executed on the blockchain environment.
Bytecode encapsulates the actions that a smart contract can perform. It contains statements and necessary information to execute the contract's functionalities. Bytecode is commonly derived from Solidity or other languages used in smart contract development. When deployed on Ethereum, bytecode is categorized into two types: creation bytecode and runtime bytecode.
\begin{enumerate}
    \item \textbf{Creation Bytecode:} The creation bytecode runs only once during the deployment of the smart contract onto the system. It is responsible for initializing the contract's initial state, including initializing variables and constructor functions. Creation bytecode does not reside within the deployed smart contract on the blockchain network.
    \item \textbf{Runtime Bytecode:} Runtime bytecode contains executable information about the smart contract and is deployed onto the blockchain network.
\end{enumerate}
Once a smart contract has been compiled into bytecode, it can be deployed onto the blockchain and executed by nodes within the network. Nodes execute the bytecode's statements to determine the behavior and interactions of the smart contract.

Bytecode is highly deterministic and remains immutable after compilation. It provides participants in the blockchain network the ability to inspect and verify smart contracts before deployment.

In summary, bytecode serves as a bridge between high-level programming languages and the blockchain environment, enabling smart contracts to be deployed and executed. Its deterministic nature and pre-deployment verifiability contribute to the security and reliability of smart contract implementations.

\subsection{Opcode of Smart Contracts}

Opcode in smart contracts refers to the executable machine instructions used in a blockchain environment to perform the functions of the smart contract. Opcodes are low-level machine commands used to control the execution process of the contract on a blockchain virtual machine, such as the Ethereum Virtual Machine (EVM).

Each opcode represents a specific task within the smart contract, including logical operations, arithmetic calculations, memory management, data access, calling and interacting with other contracts in the Blockchain network, and various other tasks. Opcodes define the actions that a smart contract can perform and specify how the contract's data and state are processed. These opcodes are listed and defined in the bytecode representation of the smart contract.

The use of opcodes provides flexibility and standardization in implementing the functionalities of smart contracts. Opcodes ensure consistency and security during the execution of the contract on the blockchain, and play a significant role in determining the behavior and logic of the smart contract.

\subsection{Control Flow Graph}

CFG is a powerful data structure in the analysis of Solidity source code, used to understand and optimize the control flow of a program extracted from the bytecode of a smart contract. The CFG helps determine the structure and interactions between code blocks in the program, providing crucial information about how the program executes and links elements in the control flow.

Specifically, CFG identifies jump points and conditions in Solidity bytecode to construct a control flow graph. This graph describes the basic blocks and control branches in the program, thereby creating a clear understanding of the structure of the Solidity program. With CFG, we can identify potential issues in the program such as infinite loops, incorrect conditions, or security vulnerabilities. By examining control flow paths in CFG, we can detect logic errors or potential unwanted situations in the Solidity program.

Furthermore, CFG supports the optimization of Solidity source code. By analyzing and understanding the control flow structure, we can propose performance and correctness improvements for the Solidity program. This is particularly crucial in the development of smart contracts on the Ethereum platform, where performance and security play essential roles.

In conclusion, CFG is a powerful representation that allows us to analyze, understand, and optimize the control flow in Solidity programs. By constructing control flow graphs and analyzing the control flow structure, we can identify errors, verify correctness, and optimize Solidity source code to ensure performance and security.

\section{Related work} \label{related_work}

This section will review existing works on smart contract vulnerability detection, including conventional methods, single learning model and multimodal learning approaches.

\subsection{Static and dynamic method}

There are many efforts in vulnerability detection in smart contracts through both static and dynamic analysis. These techniques are essential for scrutinizing both the source code and the execution process of smart contracts to uncover syntax and logic errors, including assessments of input variable validity and string length constraints. Dynamic analysis evaluates the control flow during smart contract execution, aiming to unearth potential security flaws. In contrast, static analysis employs approaches such as symbolic execution and tainting analysis. Taint analysis, specifically, identifies instances of injection vulnerabilities within the source code.

Recent research studies have prioritized control flow analysis as the primary approach for smart contract vulnerability detection. Notably, Kushwaha et al. \cite{Tool_review_SC} have compiled an array of tools that harness both static analysis techniques—such as those involving source code and bytecode—and dynamic analysis techniques via control flow scrutiny during contract execution. A prominent example of static analysis is Oyente \cite{oyente}, a tool dedicated to smart contract examination. Oyente employs control flow analysis and static checks to detect vulnerabilities like Reentrancy attacks, faulty token issuance, integer overflows, and authentication errors. Similarly, Slither \cite{slither}, a dynamic analysis tool, utilizes control flow analysis during execution to pinpoint security vulnerabilities, encompassing Reentrancy attacks, Token Issuance Bugs, Integer Overflows, and Authentication Errors. It also adeptly identifies concerns like Transaction Order Dependence (TOD) and Time Dependence.

Beyond static and dynamic analysis, another approach involves fuzzy testing. In this technique, input strings are generated randomly or algorithmically to feed into smart contracts, and their outcomes are verified for anomalies. Both Contract Fuzzer \cite{contractFuzzer} and xFuzz \cite{xFuzz} pioneer the use of fuzzing for smart contract vulnerability detection. Contract Fuzzer employs concolic testing, a hybrid of dynamic and static analysis, to generate test cases. Meanwhile, xFuzz leverages a genetic algorithm to devise random test cases, subsequently applying them to smart contracts for vulnerability assessment.

Moreover, symbolic execution stands as an additional method for in-depth analysis. By executing control flow paths, symbolic execution allows the generation of generalized input values, addressing challenges associated with randomness in fuzzing approaches. This approach holds potential for overcoming limitations and intricacies tied to the creation of arbitrary input values.

However, the aforementioned methods often have low accuracy and are not flexible between vulnerabilities as they rely on expert knowledge, fixed patterns, and are time-consuming and costly to implement. They also have limitations such as only detecting pre-defined fixed vulnerabilities and lacking the ability to detect new vulnerabilities.

\subsection{Machine Learning method}

ML methods often use features extracted from smart contracts and employ supervised learning models to detect vulnerabilities. Recent research has indicated that research groups primarily rely on supervised learning. The common approaches usually utilize feature extraction methods to obtain CFG and Abstract Syntax Tree (AST) through dynamic and static analysis tools on source code or bytecode. Th
ese studies \cite{GCN, SC_GNN} used a sequential model of Graph Neural Network to process opcodes and employed LSTM to handle the source code. Besides, a team led by Nguyen Hoang has developed Mando Guru \cite{MandoGuru}, a GNN model to detect vulnerabilities in smart contracts. Their team applied additional methods such as Heterogeneous Graph Neural Network, Coarse-Grained Detection, and Fine-Grained Detection. They leveraged the control flow graph (CFG) and call graph (CG) of the smart contract to detect 7 vulnerabilities. Their approach is capable of detecting multiple vulnerabilities in a single smart contract. The results are represented as nodes and paths in the graph. Additionally, Zhang Lejun et al. \cite{anovelSC} also utilized ensemble learning to develop a 7-layer convolutional model that combined various neural network models such as CNN, RNN, RCN, DNN, GRU, Bi-GRU, and Transformer. Each model was assigned a different role in each layer of the model.

\subsection{Multimodal Learning}
 The HyMo Framework \cite{hymo}, introduced by Chen et al. in 2020, presented a multimodal deep learning model used for smart contract vulnerability detection illustrates the components of the HyMo Framework. This framework utilizes two attributes of smart contracts, including source code and opcodes. After preprocessing these attributes, the HyMo framework employs FastText for word embedding and utilizes two Bi-GRU models to extract features from these two attributes.
 
Another framework, the HYDRA framework, proposed by Chen and colleagues \cite{hydra}, utilizes three attributes, including API, bytecode, and opcode as input for three branches in the multimodal model to classify malicious software. Each branch processes the attributes using basic neural networks, and then the outputs of these branches are connected through fully connected layers and finally passed through the Softmax function to obtain the final result.

And most recently, Jie Wanqing and colleagues have published a study \cite{2023novel} utilizing four attributes of smart contracts (SC), including source code, Static Single Assigment (SSA), CFG, and bytecode. With these four attributes, they construct three layers: SC, BB, and EVMB. Among these, the SC layer employs source code for attribute extraction using Word2Vec and BERT, the BB layer uses SSA and CFG generated from the source code, and finally, the EVMB layer employs assembly code and CFG derived from bytecode. Additionally, the authors combine these classes through various methods and undergo several distinct steps.

These models yield promising results in terms of Accuracy, with HyMo \cite{hymo} achieving approximately 0.79\%, HYDRA \cite{hydra} surpassing it with around 0.98\% and the multimodal AI of Jie et al. \cite{2023novel}. achieved high-performance results ranging from 0.94\% to 0.99\% across various test cases. With these achieved results, these studies have demonstrated the power of multimodal models compared to unimodal models in classifying objects with multiple attributes. However, the limitations within the scope of the paper are more constrained by implementation than design choices. They utilized word2vec that lacks support for out-of-vocabulary words. To address this constraint, they proposed substituting word2vec with the fastText NLP model. Subsequently, their vulnerability detection framework was modeled as a binary classification problem within a supervised learning paradigm. In this work, their primary focus was on determining whether a contract contains a vulnerability or not. A subsequent task could delve into investigating specific vulnerability types through desired multi-class classification.

From the evaluations presented in this section, we have identified the strengths and limitations of existing literature. It is evident that previous works have not fully optimized the utilization of Smart Contract data and lack the incorporation of a diverse range of deep learning models. While unimodal approaches have not adequately explored data diversity, multimodal ones have traded-off construction time for classification focus, solely determining whether a smart contract is vulnerable or not.

In light of these insights, we propose a novel framework that leverages the advantages of three distinct deep learning models including BERT, GNN, and BiLSTM. Each model forms a separate branch, contributing to the creation of a unified architecture. Our approach adopts a multi-class classification task, aiming to collectively improve the effectiveness and diversity of vulnerability detection. By synergistically integrating these models, we strive to overcome the limitations of the existing literature and provide a more comprehensive solution.

\section{Methodology} \label{methodology}
This section provides the outline of our proposed approach for vulnerability detection in smart contracts. Additionally, by employing multimodal learning, we generate a comprehensive view of the smart contract, which allows us to represent of smart contract with more relevant features and boost the effectiveness of the vulnerability detection model.

\begin{figure*}[h]
\centering
\includegraphics[width=0.9\textwidth]{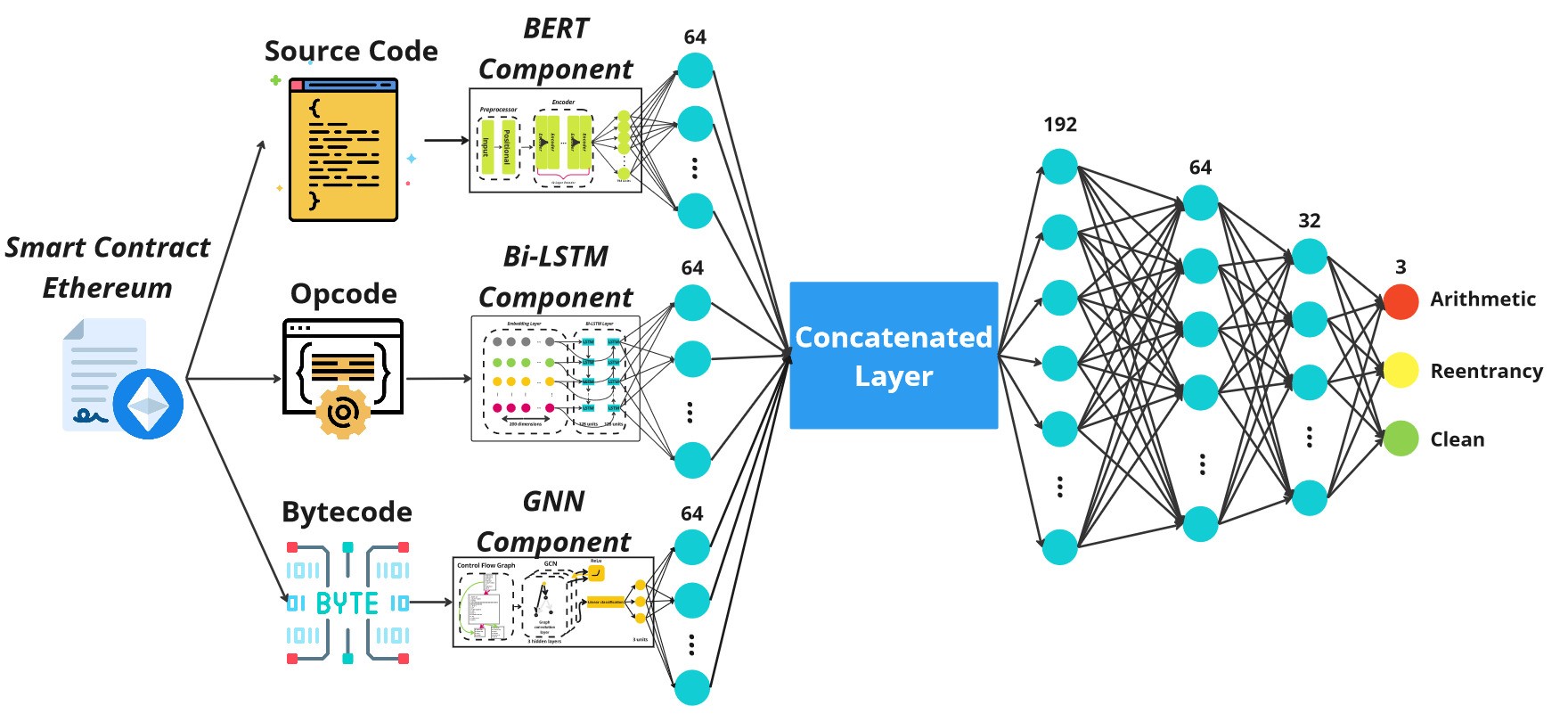}
\caption{The overview of VulnSense framework.} \label{generalmodel}
\end{figure*}
 
\subsection{An overview of architecture}
Our proposed approach, VulnSense, is constructed upon a multimodal deep learning framework consisting of three branches, including BERT, BiLSTM, and GNN, as illustrated in \textbf{Figure \ref{generalmodel}}. More specifically, the first branch is the BERT model, which is built upon the Transformer architecture and employed to process the source code of the smart contract. Secondly, to handle and analyze the opcode context, the BiLSTM model is applied in the second branch. Lastly, the GNN model is utilized for representing the CFG of bytecode in the smart contract.

This integrative methodology leverages the strengths of each component to comprehensively assess potential vulnerabilities within smart contracts. The fusion of linguistic, sequential, and structural information allows for a more thorough and insightful evaluation, thereby fortifying the security assessment process. This approach presents a robust foundation for identifying vulnerabilities in smart contracts and holds promise for significantly reducing risks in blockchain ecosystems.

\subsection{Bidirectional Encoder Representations from Transformers (BERT)}
\begin{figure}[h] 
\centering
\includegraphics[width=0.4\textwidth]{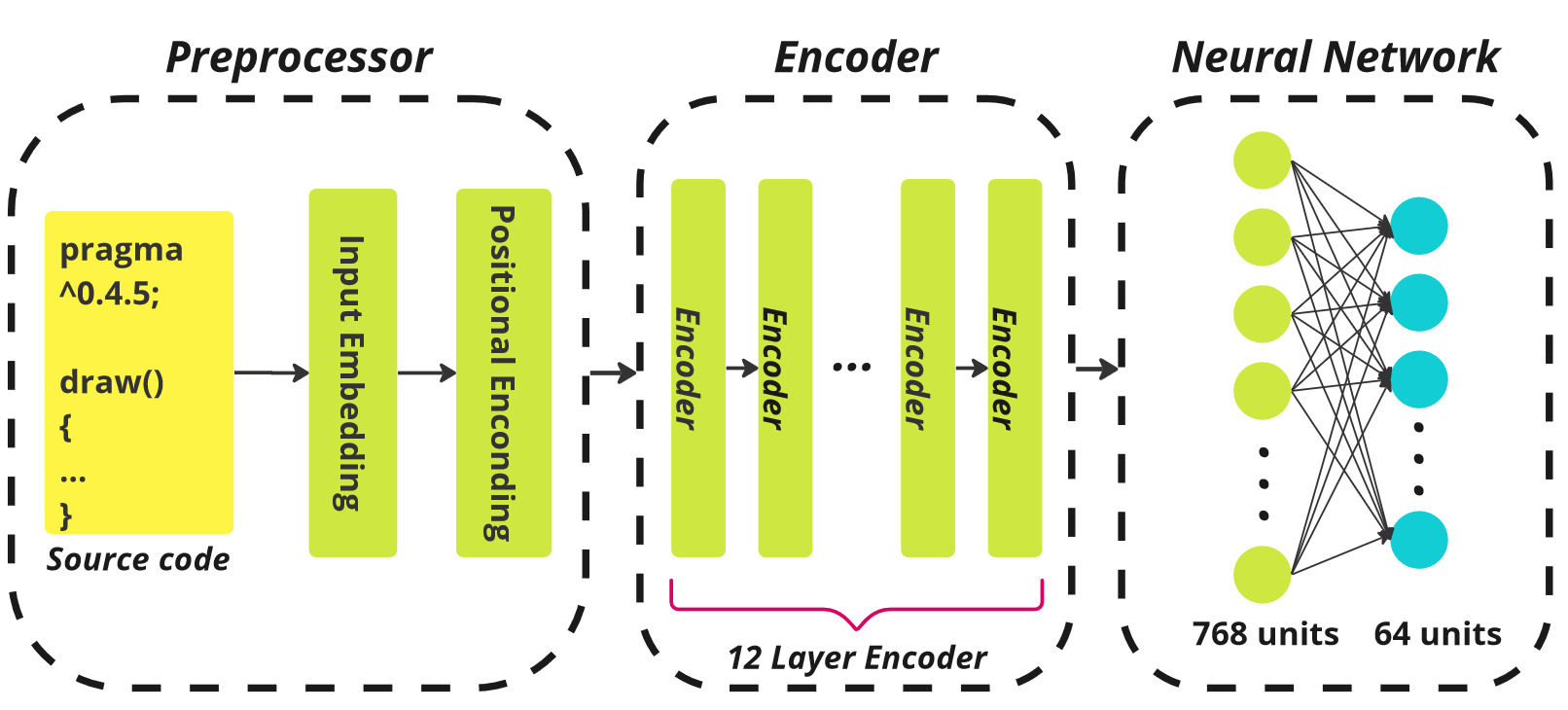}
\caption{The architecture of BERT component in VulnSense} \label{fig:bert_model}
\end{figure}

In this study, to capture high-level semantic features from the source code and enable a more in-depth understanding of its functionality, we designed a BERT network which is a branch in our Multimodal.

As in \textbf{Figure \ref{fig:bert_model}}, BERT model consists of 3 blocks: Preprocessor, Encoder and Neural network. More specifically, the Preprocessor processes the inputs, which is the source code of smart contracts. The inputs are transformed into vectors through the input embedding layer, and then pass through the $positional\_encoding$  layer to add positional information to the words. Then, the $\mathbf{preprocessed}$ values are fed into the encoding block  to compute relationships between words. The entire encoding block consists of 12 identical encoding layers stacked on top of each other. Each encoding layer comprises two main parts: a self-attention layer and a feed-forward neural network. The output $\mathbf{encoded}$ forms a vector space of length 768. Subsequently, the $\mathbf{encoded}$ values are passed through a simple neural network. The resulting $\mathbf{bert\_output}$ values constitute the output of this branch in the multimodal model. Thus, the whole BERT component could be demonstrated as follows:
\begin{equation} 
    \mathbf{preprocessed} = positional\_encoding(\mathbf{e}(input))
    \label{preprocessed}
\end{equation}
\begin{equation}
    \mathbf{encoded} = Encoder(\mathbf{preprocessed})
    \label{encoded}
\end{equation}
\begin{equation}
    \mathbf{bert\_output} = NN(\mathbf{encoded})
    \label{nn}
\end{equation}
where, (\ref{preprocessed}), (\ref{encoded}) and (\ref{nn}) represent the Preprocessor block, Encoder block and Neural Network block, respectively.

\subsection{Bidirectional long-short term memory (BiLSTM)}
\label{subsec_lstm}
\begin{figure}[h]
\centering
\includegraphics[width=0.4\textwidth]{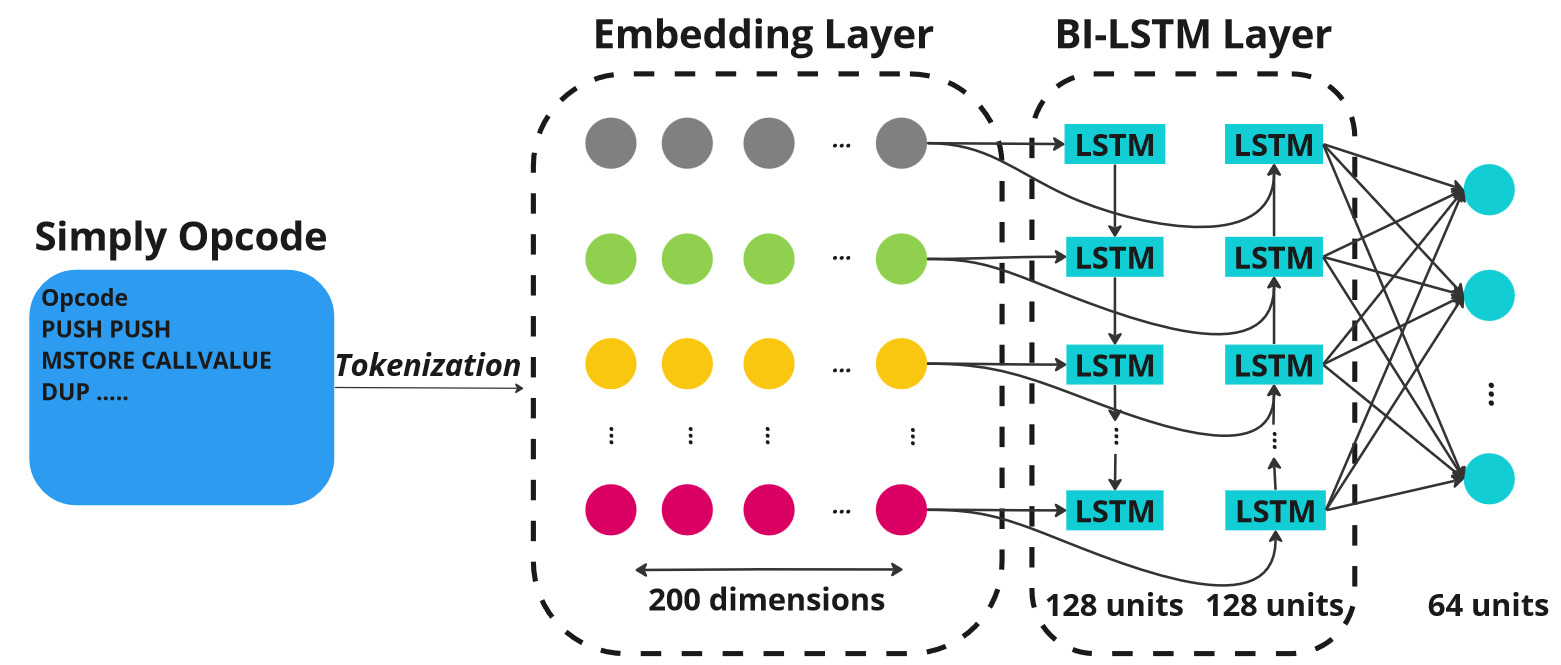}
\caption{The architecture of BiLSTM component in VulnSense.} \label{bilstm}
\end{figure}
Toward the opcode, we applied the BiLSTM which is another branch of our Multimodal approach to analysis the contextual relation of opcodes and contribute crucial insights into the code's execution flow. By processing Opcode sequentially, we aimed to capture potential vulnerabilities that might be overlooked by solely considering structural information. 

In detail, as in \textbf{Figure \ref{bilstm}}, we first tokenize the opcodes and convert them into integer values. The opcode features tokenized are embedded into a dense vector space using an $embedding$ layer which has 200 dimensions.
\begin{equation}\label{embedded}
\begin{split}
    \mathbf{token} = Tokenize(\mathbf{opcode}) \\
    \mathbf{vector\_space} = Embedding(\mathbf{token})
\end{split}
\end{equation}

Then, the opcode vector is fed into two BiLSTM layers with 128 and 64 units respectively. Moreover, to reduce overfitting, the Dropout layer is applied after the first BiLSTM layer  as in (\ref{bilstm1}).
\begin{equation}\label{bilstm1}
\begin{split}
    \mathbf{bi\_lstm1} = Bi\_LSTM(128)(\mathbf{vector\_space}) \\
    \mathbf{r} = Dropout(dense(\mathbf{bi\_lstm1})) \\
    \mathbf{bi\_lstm2} = Bi\_LSTM(128)(\mathbf{r})
\end{split}
\end{equation}

Finally, the output of the last BiLSTM layer is then fed into a dense layer with 64 units and ReLU activation function as in (\ref{bilstm2}).
\begin{equation}\label{bilstm2}
\begin{split}
    \mathbf{lstm\_output} = Dense(64, relu)(\mathbf{bi\_lstm2})
\end{split}
\end{equation}

\subsection{Graph Neural Network (GNN)}
\begin{figure}[hpt]
\centering
\includegraphics[width=0.4\textwidth]{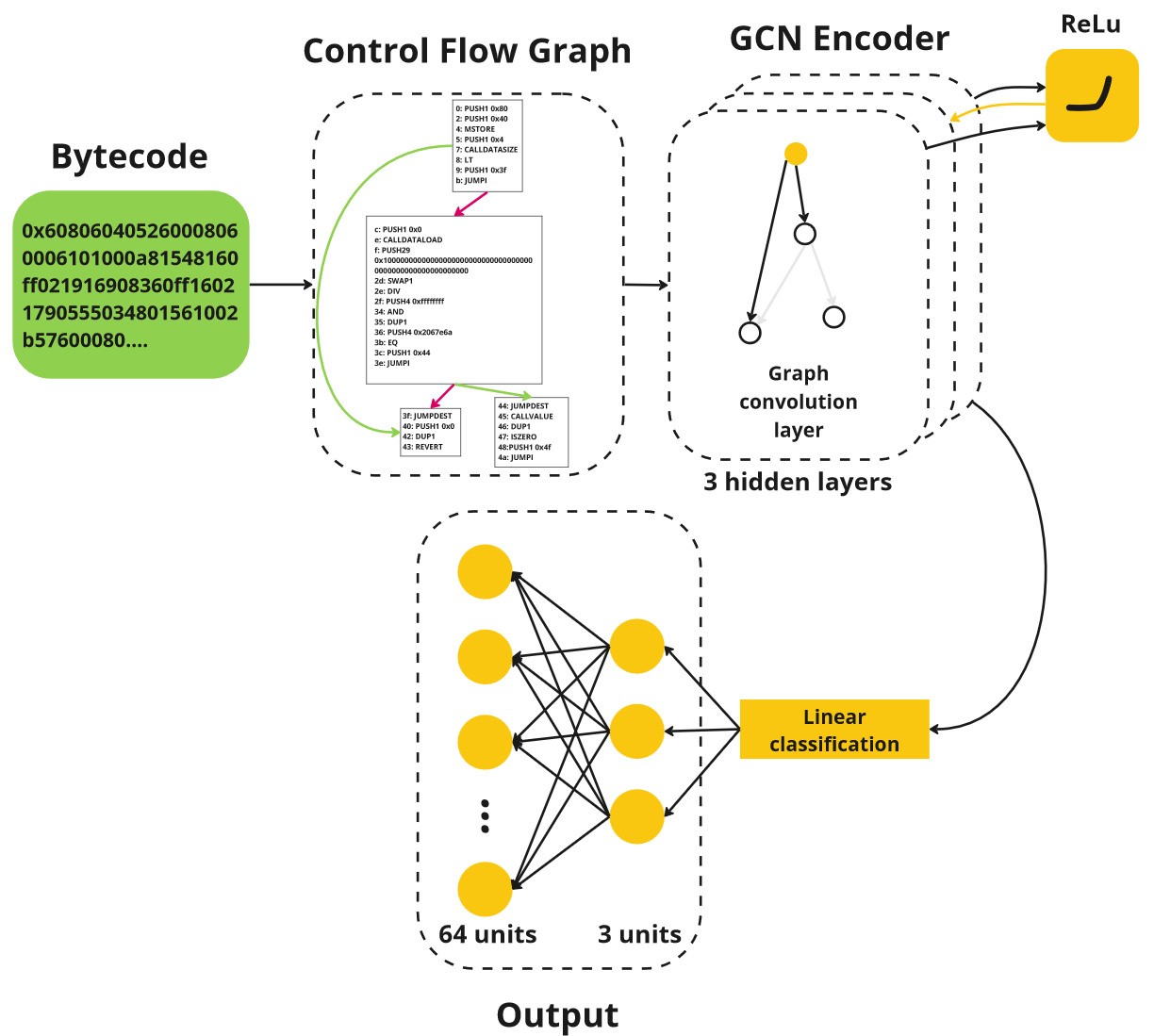}
\caption{The architecture of GNN component in VulnSense} \label{fig:gnncomponent}
\end{figure}

To offer insights into the structural characteristics of smart contracts based on bytecode, we present a CFG-based GNN model which is the third branch of our multimodal model, as shown in \textbf{Figure \ref{fig:gnncomponent}}. 

In this branch, we firstly extract the CFG from the bytecode, and then use OpenAI's embedding API to encode the nodes and edges of the CFG into vectors, as in (\ref{encodeOpenAI}).
\begin{equation}
    \mathbf{encode} = Encoder(edges, nodes)
    \label{encodeOpenAI}
\end{equation}

The encoded vectors have a length of 1536. These vectors are then passed through 3 GCN layers with ReLU activation functions (\ref{GCNencode}), with the first layer having an input length of 1536 and an output length of a custom hidden\_channels ($hc$) variable.

\begin{equation}
\begin{split}
    \mathbf{GCN1} = GCNConv(1536, relu)(\mathbf{encode})
    \\
    \mathbf{GCN2} = GCNConv(hc, relu)(\mathbf{GCN1})
    \\
    \mathbf{GCN3} = GCNConv(hc)(\mathbf{GCN2})
    \label{GCNencode}
\end{split}
\end{equation}



Finally, to feed into the multimodal deep learning model, the output of the GCN layers is fed into 2 dense layers with 3 and 64 units respecitvely, as described in (\ref{outGNN}). 
\begin{equation}
\begin{split}
    \mathbf{d1\_gnn} = Dense(3, relu)(\mathbf{GCN3})\\
    \mathbf{gnn\_output} = Dense(64, relu)(\mathbf{d1\_gnn})
    \label{outGNN}
\end{split}
\end{equation}

\subsection{Multimodal}
Each of these branches contributes a unique dimension of analysis, allowing us to capture intricate patterns and nuances present in the smart contract data. Therefore, we adopt an innovative approach by synergistically concatinating the outputs of three distinctive models including BERT $\mathbf{bert\_output}$ (\ref{nn}), BiLSTM $\mathbf{lstm\_output}$ (\ref{bilstm2}), and GNN $\mathbf{gnn\_output}$ (\ref{outGNN}) to enhance the accuracy and depth of our predictive model, as shown in (\ref{Concatenate}):
\begin{equation}
\begin{split}
   \mathbf{c} = \mathrm{Concatenate}([\mathbf{bert\_output}, \\\mathbf{lstm\_output}, \mathbf{gnn\_output}])
    \label{Concatenate}
\end{split}
\end{equation}

Then the output $\mathbf{c}$ is transformed into a 3D tensor with dimensions (batch\_size, 194, 1) using the Reshape layer (\ref{Reshape}):
\begin{equation}
    \mathbf{c\_reshaped} = \mathrm{Reshape}((194, 1))(\mathbf{c})\label{Reshape}
\end{equation}

Next, the transformed tensor $\mathbf{c\_reshaped}$ is passed through a 1D convolutional layer (\ref{Conv1D}) with 64 filters and a kernel size of 3, utilizing the rectified linear activation function:
\begin{equation}
    \mathbf{conv\_out} = \mathrm{Conv1D}(64, 3, \mathrm{relu})(\mathbf{c\_reshaped}) \label{Conv1D}
\end{equation}

The output from the convolutional layer is then flattened (\ref{Flatten}) to generate a 1D vector:
\begin{equation}
    \mathbf{f\_out} = \mathrm{Flatten}()(\mathbf{conv\_out}) \label{Flatten}
\end{equation}

The flattened tensor $\mathbf{f\_out}$ is subsequently passed through a fully connected layer with length 32 and an adjusted rectified linear activation function as in (\ref{fcn}):
\begin{equation}
 \mathbf{d\_out} = \mathrm{Dense}(32, \mathrm{relu})(\mathbf{f\_out}) \label{fcn}
\end{equation}

Finally, the output is passed through the softmax activation function (\ref{sMax}) to generate a probability distribution across the three output classes:
\begin{equation}
    \mathbf{\widetilde{y}} = \mathrm{Dense}(3, \mathrm{softmax})(\mathbf{d\_out}) \label{sMax}
\end{equation}

This architecture forms the final stages of our model, culminating in the generation of predicted probabilities for the three output classes. 

\section{Experiments and Analysis} \label{experiments}

\subsection{Experimental Settings and Implementation}
In this work, we utilize a virtual machine (VM) of Intel Xeon(R) CPU E5-2660 v4 @ 2.00GHz x 24, 128 GB of RAM, and Ubuntu 20.04 version for our implementation. Furthermore, all experiments are evaluated under the same experimental conditions. The proposed model is implemented using Python programming language and utilized well-established libraries such as TensorFlow, Keras.

For all the experiments, we have utilized the fine-tune strategy to improve the performance of these models during the training stage. We set the batch size as 32 and the learning rate of optimizer Adam with 0.001. Additionally, to escape overfitting data, the dropout operation ratio has been set to 0.03.


\subsection{Performance Metrics}

We evaluate our proposed method via 4 following metrics, including Accuracy, Precision, Recall, F1-Score. Since our work conducts experiments in multi-classes classification tasks, the value of each metric is computed based on a 2D confusion matrix which includes True Positive (TP), True Negative (TN), False Positive (FP) and False Negative (FN).

\textit{Accuracy} is the ratio of correct predictions $TP,~TN$ over all predictions. 


\textit{Precision} measures the proportion of $TP$ over all samples classified as positive. 

\textit{Recall} is defined the proportion of $TP$ over all positive instances in a testing dataset.

\textit{F1-Score} is the Harmonic Mean of $Precision$ and $Recall$.

\subsection{Dataset and Preprocessing} \label{subsec:dataset}
\begin{table}[h]
    \caption{Distribution of Labels in the Dataset}\label{tab:dataset}
    \begin{center}
    \begin{tabular}{|c|c|}
      \hline Vulnerability Type  & Contracts \\  \hline
      Arithmetic & 631\\
      Re-entrancy & 591\\
      Non-Vulnerability & 547\\
      \hline
    \end{tabular}
    \end{center} 
\end{table}
In this dataset, we combine three datasets, including Smartbugs Curated \cite{durieux2020empirical, ferreira2020smartbugs}, SolidiFI-Benchmark \cite{ghaleb2020effective}, and Smartbugs Wild \cite{durieux2020empirical, ferreira2020smartbugs}. For the Smartbugs Wild dataset, we collect smart contracts containing a single vulnerability (either an Arithmetic vulnerability or a Reentrancy vulnerability). The identification of vulnerable smart contracts is confirmed by at least two vulnerability detection tools currently available. In total, our dataset includes 547 Non-Vulnerability, 631 Arithmetic Vulnerabilities of Smart Contracts, and 591 Reentrancy Vulnerabilities of Smart Contracts, as shown in \textbf{Table \ref{tab:dataset}}.

\subsubsection{Source Code Smart Contract}
When programming, developers often have the habit of writing comments to explain their source code, aiding both themselves and other programmers in understanding the code snippets. BERT, a natural language processing model, takes the source code of smart contracts as its input. From the source code of smart contracts, BERT calculates the relevance of words within the code. Comments present within the code can introduce noise to the BERT model, causing it to compute unnecessary information about the smart contract's source code. Hence, preprocessing of the source code before feeding it into the BERT model is necessary.

\begin{figure}[h]
\centering
\includegraphics[width=0.4\textwidth]{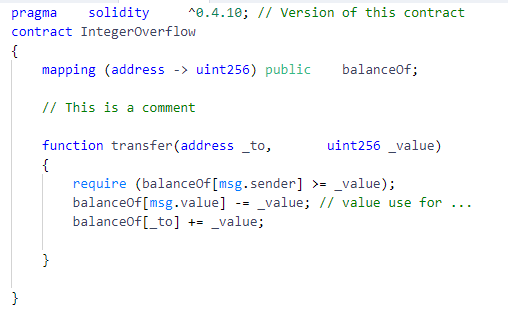}
\caption{An example of Smart Contract Prior to Processing} \label{code:trashSC}
\end{figure}

\begin{figure}[h]
\centering
\includegraphics[width=0.4\textwidth]{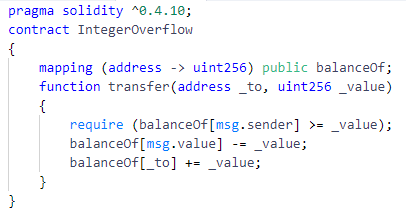}
\caption{An example of processed Smart Contract} \label{code:cleanSC}
\end{figure}

Moreover, removing comments from the source code also helps reduce the length of the input when fed into the model. To further reduce the source code length, we also eliminate extra blank lines and unnecessary whitespace. \textbf{Figure \ref{code:trashSC}} provides an example of an unprocessed smart contract from our dataset. This contract contains comments following the '//' syntax, blank lines, and excessive white spaces that do not adhere to programming standards. \textbf{Figure \ref{code:cleanSC}} represents the smart contract after undergoing processing.

\subsubsection{Opcode Smart Contracts}
We proceed with bytecode extraction from the source code of the smart contract, followed by opcode extraction through the bytecode. The opcodes within the contract are categorized into 10 functional groups, totaling 135 opcodes, according to the Ethereum Yellow Paper \cite{yellow_paper}. However, we have condensed them based on \textbf{Table \ref{tab:OpcodeSimplify}}.
\begin{table}[h]
    \caption{The simplified opcode methods}\label{tab:OpcodeSimplify}
    \begin{center}
    \resizebox{0.3\textwidth}{!}{
    \begin{tabular}{|c|c|}
      \hline Substituted Opcodes  & Original Opcodes \\  \hline
      DUP & DUP1-DUP16\\
      SWAP & SWAP1-SWAP16\\
      PUSH & PUSH5-PUSH32\\
      LOG & LOG1-LOG4 \\
      \hline
    \end{tabular}}
    \end{center}
\end{table}

During the preprocessing phase, unnecessary hexadecimal characters were removed from the opcodes. The purpose of this preprocessing is to utilize the opcodes for vulnerability detection in smart contracts using the BiLSTM model.

In addition to opcode preprocessing, we also performed other preprocessing steps to prepare the data for the BiLSTM model. Firstly, we tokenized the opcodes into sequences of integers. Subsequently, we applied padding to create opcode sequences of the same length. The maximum length of opcode sequences was set to 200, which is the maximum length that the BiLSTM model can handle.

After the padding step, we employ a Word Embedding layer to transform the encoded opcode sequences into fixed-size vectors, serving as inputs for the BiLSTM model. This enables the BiLSTM model to better learn the representations of opcode sequences.

In general, the preprocessing steps we performed are crucial in preparing the data for the BiLSTM model and enhancing its performance in detecting vulnerabilities in Smart Contracts.

\subsubsection{Control Flow Graph}
\begin{figure}[h]
\centering
\includegraphics[width=0.4\textwidth]{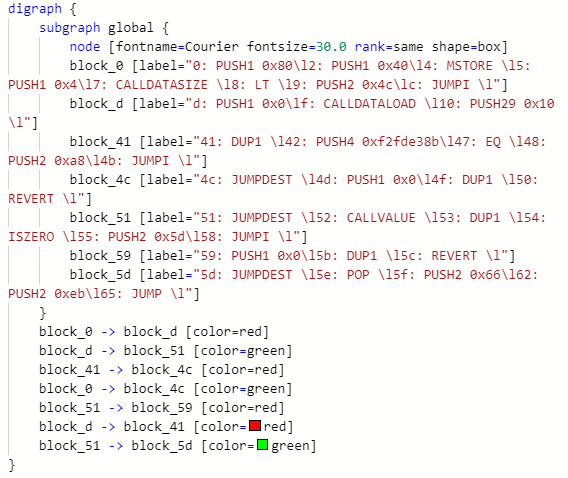}
\caption{Graph Extracted from Bytecode} \label{code:cfgextract}
\end{figure}

\begin{figure}[h]
\centering
\includegraphics[width=0.4\textwidth]{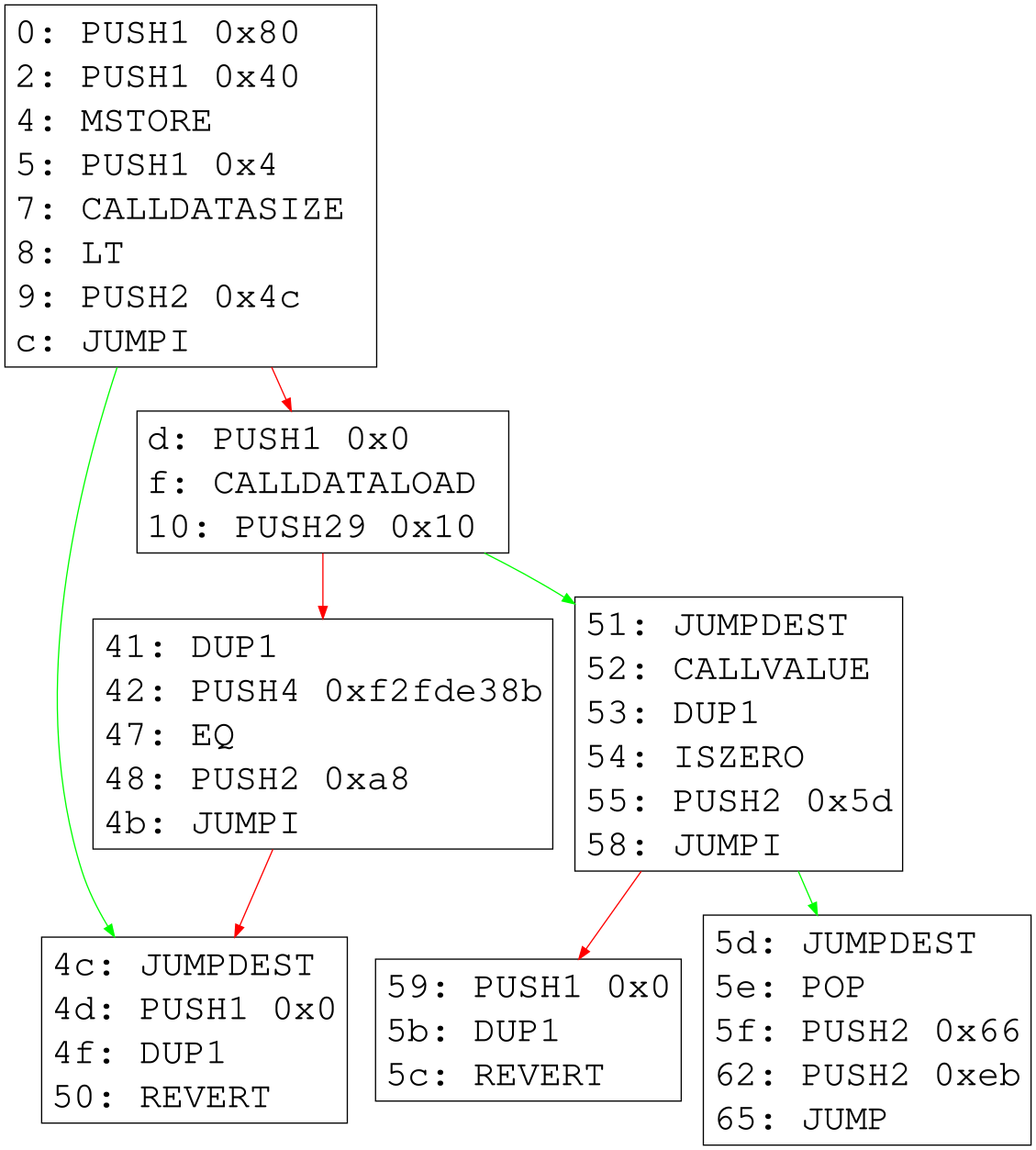}
\caption{Visualize Graph by .cgf.gv file } \label{fig:cfg_pre}
\end{figure}
First, we extract bytecode from the smart contract, then extract the CFG through bytecode into .cfg.gv files as shown in \textbf{Figure \ref{code:cfgextract}}. From this .cfg.gv file, a CFG of a Smart Contract through bytecode can be represented as shown in \textbf{Figure \ref{fig:cfg_pre}}. The nodes in the CFG typically represent code blocks or states of the contract, while the edges represent control flow connections between nodes.

To train the GNN model, we encode the nodes and edges of the CFG into numerical vectors. One approach is to use embedding techniques to represent these entities as vectors. In this case, we utilize the OpenAI API embedding to encode nodes and edges into vectors of length 1536. This could be a customized approach based on OpenAI's pre-trained deep learning models. Once the nodes and edges of the CFG are encoded into vectors, we employ them as inputs for the GNN model.

\subsection{Experimental Scenarios}


To prove the efficiency of our proposed model and compared models, we conducted training with a total of 7 models, categorized into two types: unimodal deep learning models and multi-modal deep learning models.

On the one hand, the unimodal deep learning models consisted of models within each branch of VulnSense. On the other hand, these multimodal deep learning models are pairwise combinations of three unimodal deep learning models and VulnSense, utilizing a 2-way interaction. Specifically:
\begin{itemize}
    \item \textbf{Unimodal}:
    \begin{itemize}
        \item BiLSTM
        \item BERT
        \item GNN
    \end{itemize}
    \item \textbf{Multimodal}:
    \begin{itemize}
        \item Multimodal BERT - BiLSTM (\textbf{M1})
        \item Multimodal BERT - GNN (\textbf{M2})
        \item Multimodal BiLSTM - GNN (\textbf{M3})
        \item \textbf{VulnSense} (as mentioned in \textbf{Section \ref{methodology}})
    \end{itemize}    
\end{itemize}
\begin{table*}[h]
\centering
\caption{The performance of 7 models}
\label{tab_multimodal_res}
\resizebox{0.9\textwidth}{!}{
\begin{tabular}{|c|c|c|c|c|c|c|c|c|}
\hline
\textbf{Score} & \textbf{Epoch} & \textbf{BERT} & \textbf{BiLSTM} & \textbf{GNN} & \textbf{M1} & \textbf{M2} & \textbf{M3} & \textbf{VulnSense} \\
\hline
\multirow{3}{*}{\textbf{Accuracy}} & \textbf{E10} & 0.5875 & 0.7316 & 0.5960 & 0.7429 & 0.6468 & 0.7542 & \textbf{0.7796} \\
\cline{2-9} & \textbf{E20} & 0.5903 & 0.6949 & 0.5988 & 0.7796 & 0.6553 & 0.7768 & \textbf{0.7796} \\
\cline{2-9} & \textbf{E30} & 0.6073 & 0.7146 & 0.6016 & 0.7796 & 0.6525 & 0.7683 & \textbf{0.7796} \\
\hline
\multirow{3}{*}{\textbf{Precision}} & \textbf{E10} & 0.5818 & 0.7540 & 0.4290 & 0.7749 & 0.6616 & 0.7790 & \textbf{0.7940} \\
\cline{2-9} & \textbf{E20} & 0.6000 & 0.7164 & 0.7209 & 0.7834 & 0.6800 & 0.7800 & \textbf{0.7922} \\
\cline{2-9} & \textbf{E30} & 0.6000 & 0.7329 & 0.5784 & 0.7800 & 0.7000 & 0.7700 & \textbf{0.7800} \\
\hline
\multirow{3}{*}{\textbf{Recall}} & \textbf{E10} & 0.5876 & 0.7316 & 0.5960 & 0.7429 & 0.6469 & 0.7542 & \textbf{0.7797} \\
\cline{2-9} & \textbf{E20} & 0.5900 & 0.6949 & 0.5989 & 0.7797 & 0.6600 & \textbf{0.7800} & 0.7797 \\
\cline{2-9} & \textbf{E30} & 0.6100 & 0.7147 & 0.6017 & 0.7700 & 0.6500 & 0.7700 & \textbf{0.7700} \\
\hline
\multirow{3}{*}{\textbf{F1}} & \textbf{E10} & 0.5785 & 0.7360 & 0.4969 & 0.7509 & 0.6520 & 0.7602 & \textbf{0.7830} \\
\cline{2-9} & \textbf{E20} & 0.5700 & 0.6988 & 0.5032 & 0.7809 & 0.6600 & 0.7792 & \textbf{0.7800} \\
\cline{2-9} & \textbf{E30} & 0.6000 & 0.7185 & 0.5107 & 0.7700 & 0.6500 & 0.7700 & \textbf{0.7750} \\
\hline
\end{tabular}}
\end{table*}
Furthermore, to illustrate the fast convergence and stability of multimodal method, we train and validate 7 models on 3 different mocks of training epochs including 10, 20 and 30 epochs.  

\subsection{Experimental Results}
The experimentation process for the models was carried out on the dataset as detailed in \textbf{Section \ref{subsec:dataset}}.

\subsubsection{Models performance evaluation} \label{cmpMetric}

Through the visualizations in \textbf{Table \ref{tab_multimodal_res}}, it can be intuitively answered that the ability to detect vulnerabilities in smart contracts using multi-modal deep learning models is more effective than unimodal deep learning models in these experiments. Specifically, when testing 3 multimodal models including M1, M3 and VulnSense on 3 mocks of training epochs, the results indicate that the performance is always higher than 75.09\% with 4 metrics mentioned above. Meanwhile, the testing performance of M2 and 3 unimodal models including BERT, BiLSTM and GNN are lower than 75\% with all 4 metrics. Moreover, with the testing performance on all 3 mocks of training epochs, VulnSense model has achieved the highest F1-Score with more than 77\% and Accuracy with more than 77.96\%. 

\begin{figure*}[h]
  \begin{tabular}{cccc}
    \subfloat[BERT]{%
     \includegraphics[width=0.23\textwidth]{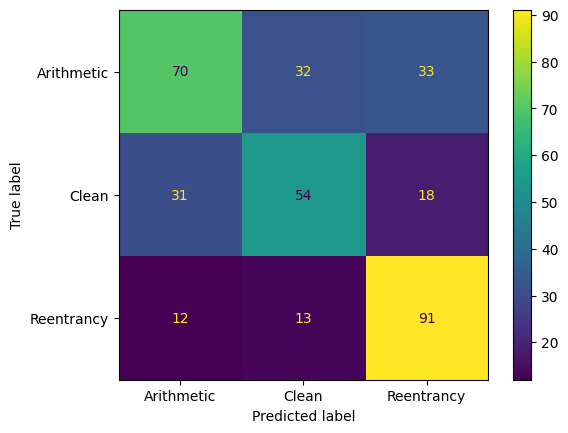}%
     } &
    \subfloat[M1]{%
     \includegraphics[width=0.23\textwidth]{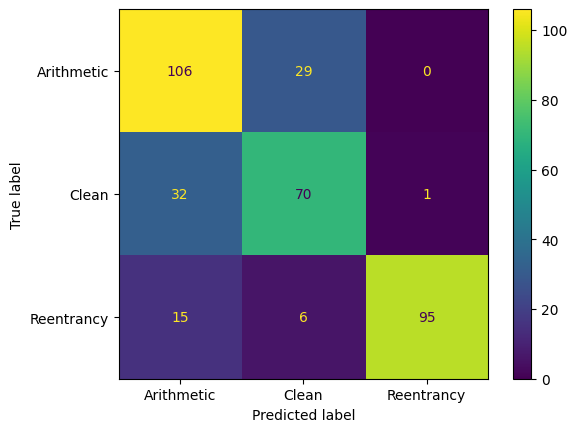}%
     } & 
     \subfloat[BiLSTM]{%
     \includegraphics[width=0.23\textwidth]{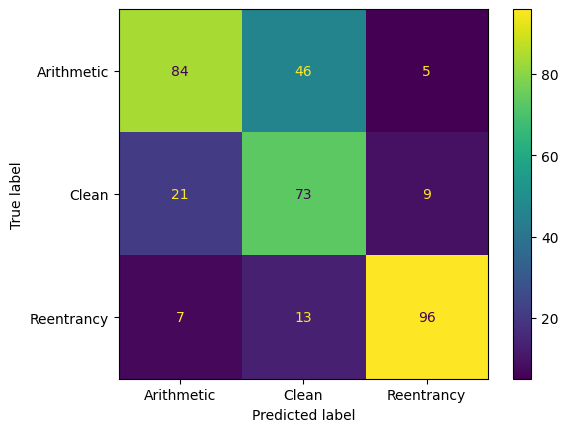}%
     } &
     \subfloat[M2]{%
     \includegraphics[width=0.23\textwidth]{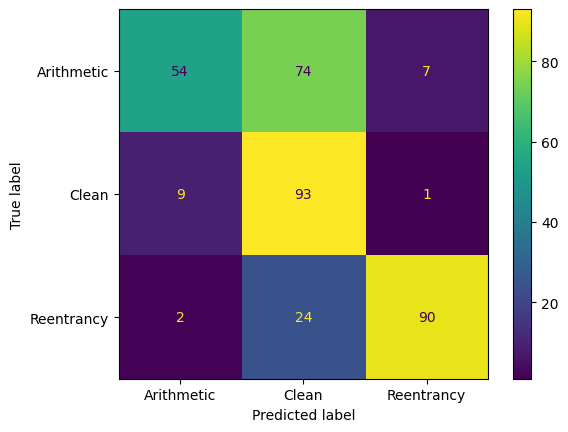}%
     }\\
    \subfloat[GNN]{%
     \includegraphics[width=0.23\textwidth]{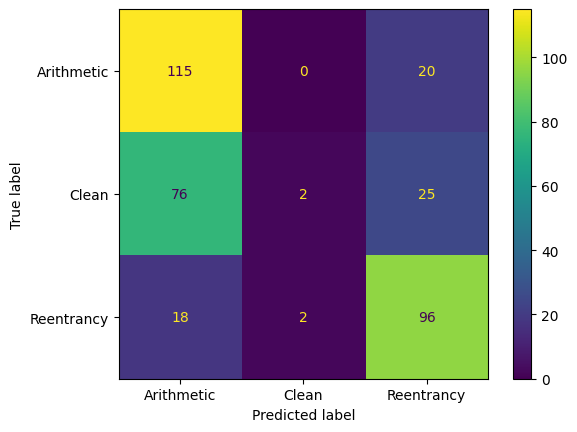}%
     }&
     \subfloat[M3]{%
     \includegraphics[width=0.23\textwidth]{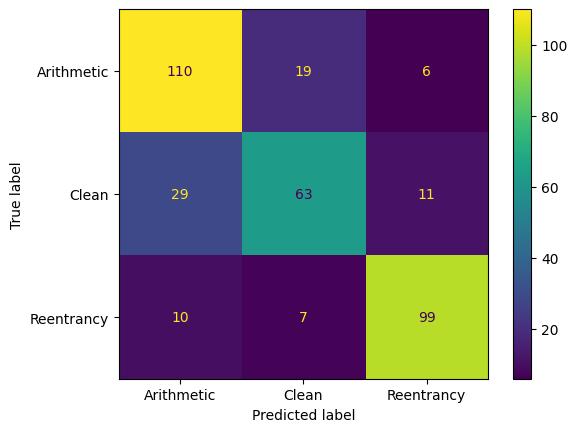}%
     }&
     \subfloat[VulnSense]{%
     \includegraphics[width=0.23\textwidth]{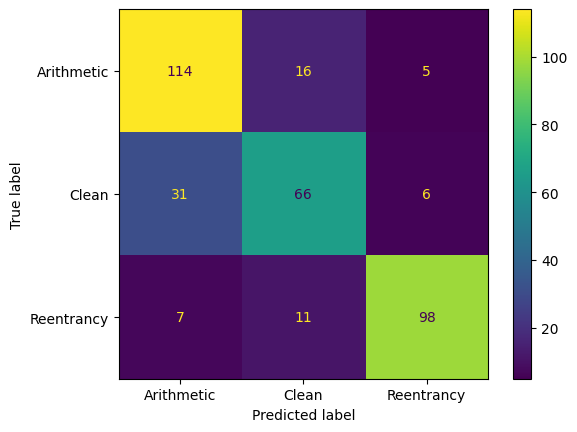}%
     }\\
  \end{tabular}
  \caption{Confusion matrices at the $30^{th}$ epoch, with (a), (c), (e) representing the unimodal models, and (b), (d), (f), (g) representing the multimodal models}
  \label{fig:c_matrixs}
\end{figure*}

In addition, \textbf{Figure \ref{fig:c_matrixs}} provides a more detailed of the performances of all 7 models at the last epoch training. It can be seen from \textbf{Figure \ref{fig:c_matrixs}} that, among these multimodal models, VulnSense performs the best, having the accuracy of Arithmetic, Reentrancy, and Clean label of 84.44\%, 64.08\% and 84.48\%, respectively, followed by M3, M1 and M2 model. Even though the GNN model, which is an unimodal model, managed to attain an accuracy rate of 85.19\% for the Arithmetic label and 82.76\% for the Reentrancy label, its performance in terms of the Clean label accuracy was merely 1.94\%. Similarity in the context of unimodal models, BiLSTM and BERT both have gived the accuracy of all 3 labels relatively low less than 80\%. 

\begin{figure*}[h]
 \centering
 \def\twidth{1}
 \subfloat[Accuracy]{%
 \includegraphics[width=0.49\textwidth]{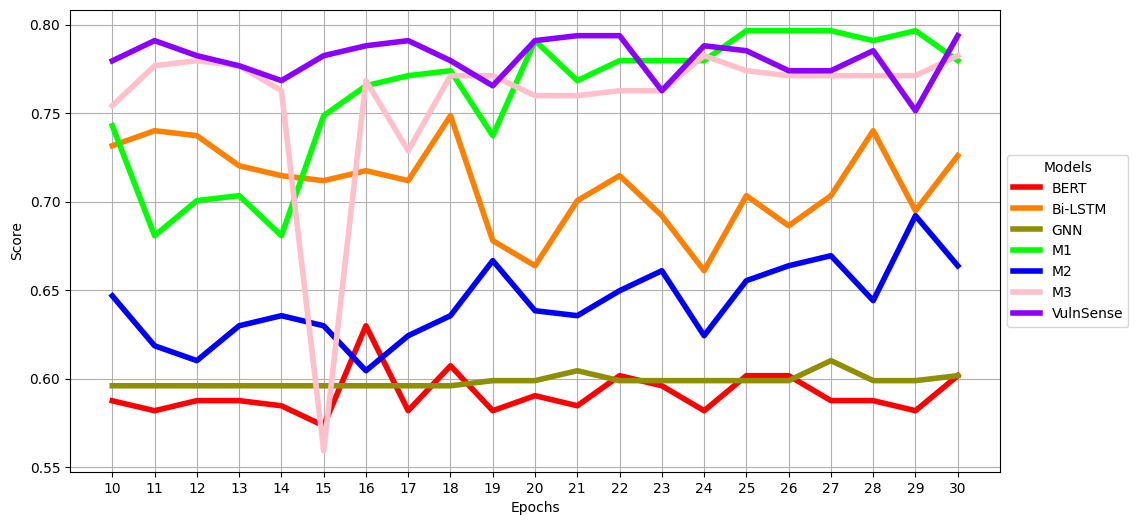}%
 } \hfill
 \subfloat[Precision]{%
 \includegraphics[width=0.49\textwidth]{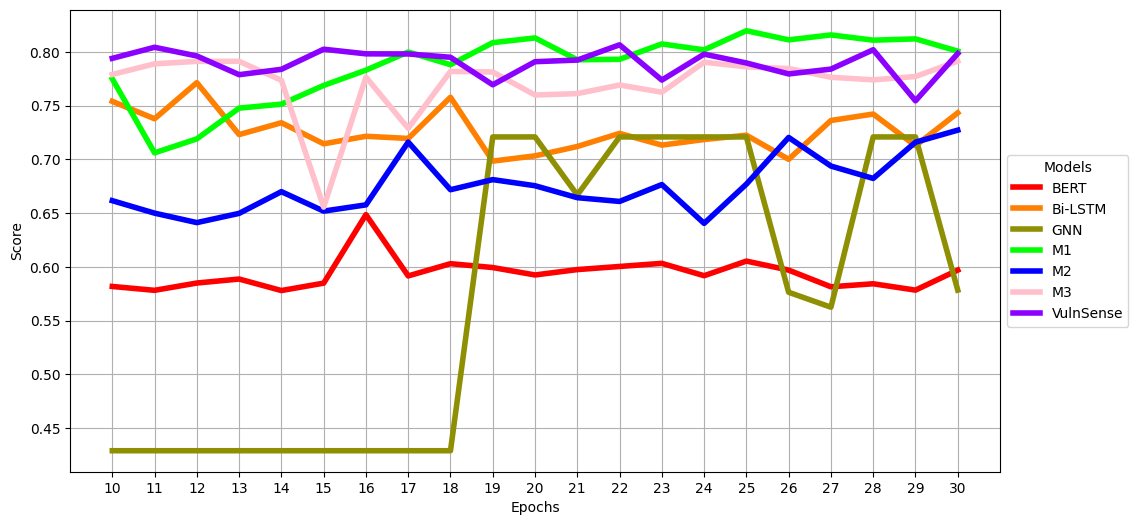}%
 } \\
 \subfloat[Recall]{%
 \includegraphics[width=0.49\textwidth]{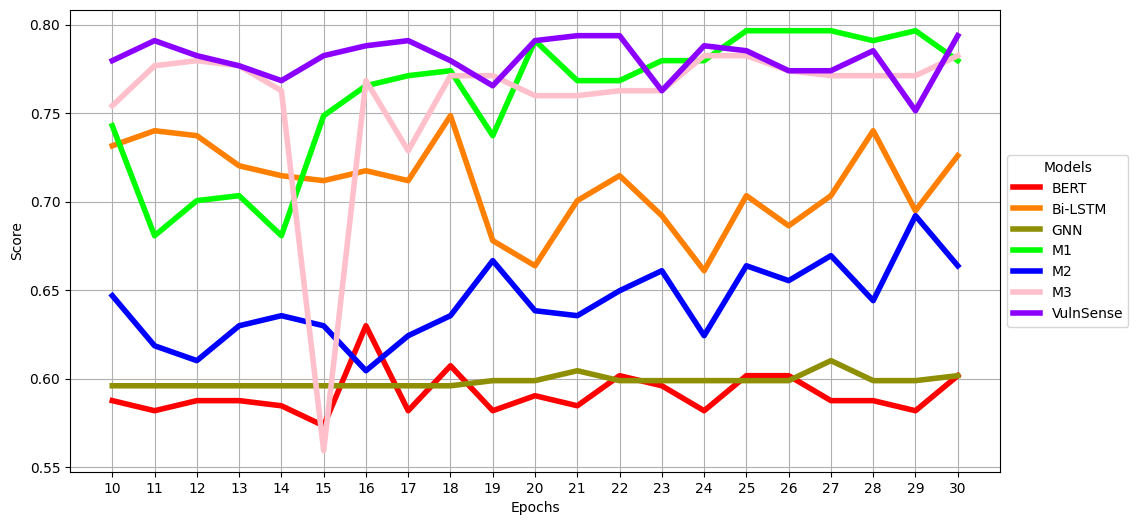}%
 } \hfill
 \subfloat[F1-Score]{%
 \includegraphics[width=0.49\textwidth]{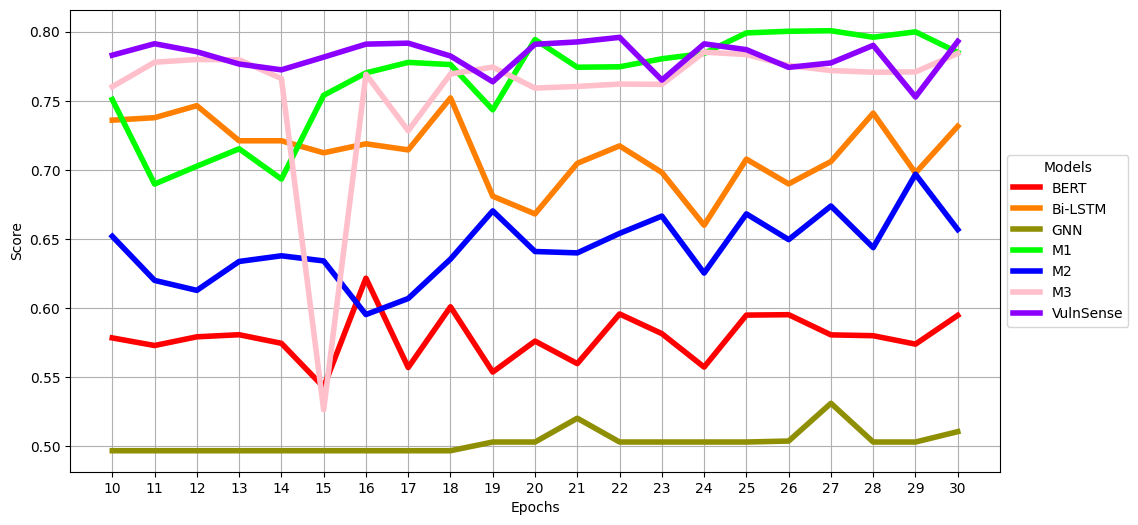}%
 }
 \caption{The performance of 7 models in 3 different mocks of training epochs.}
 \label{fig:chart_performance}
\end{figure*}

Furthermore, the results shown in \textbf{Figure \ref{fig:chart_performance}} have demonstrated the superior convergence speed and stability of VulnSense model compared to the other 6 models. In detail, through testing after 10 training epochs, VulnSense model has gained the highest performance with greater than 77.96\% in all 4 metrics. Although, VulnSense, M1 and M3 models give high performance after 30 training epochs, VulnSense model only needs to be trained for 10 epochs to achieve better convergence than the M1 and M3 models, which require 30 epochs. Besides, throughout 30 training epochs, the M3, M1, BiLSTM, and M2 models exhibited similar performance to the VulnSense model, yet they demonstrated some instability. On the one hand, the VulnSense model maintains a consistent performance level within the range of 75-79\%, on the other hand, the M3 model experienced a severe decline in both Accuracy and F1-Score values, declining by over 20\% by the $15^{th}$ epoch, indicating significant disturbance in its performance.

From this observation, these findings indicate that our proposed model, VulnSense, is more efficient in identifying vulnerabilities in smart contracts compared to these other models. Furthermore, by harnessing the advantages of multimodal over unimodal, VulnSense also exhibited consistent performance and rapid convergence.

\subsubsection{Comparisons of Time} \label{cmpTrainingTime}
\begin{figure}[h]
    \centering
    \includegraphics [width=0.4\textwidth]{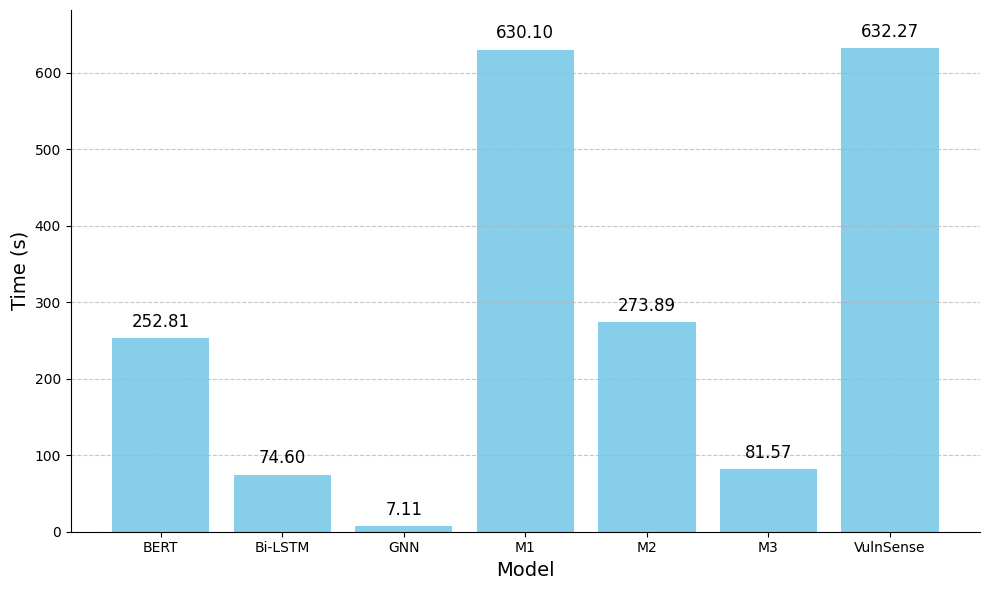}
    \caption{Comparison chart of training time for 30 epochs among models}
    \label{fig:cmpTime}
\end{figure}

\textbf{Figure \ref{fig:cmpTime}} illustrates the training time for 30 epochs of each model. 
Concerning the training time of unimodal models, on the one hand, the training time for the GNN model is very short, at only 7.114 seconds, on the other hand  BERT model reaches significantly longer training time of 252.814 seconds. For the BiLSTM model, the training time is significantly 10 times longer than that of the GNN model. Furthermore, when comparing the multimodal models, the shortest training time belongs to the M3 model (the multimodal combination of BiLSTM and GNN) at 81.567 seconds. Besides, M1, M2, and VulnSense involve the BERT model, resulting in relatively longer training times with over 270 seconds for 30 epochs. It is evident that the unimodal model significantly impacts the training time of the multimodal model it contributes to. Although VulnSense takes more time compared to the 6 other models, it only requires 10 epochs to converge. This greatly reduces the training time of VulnSense by 66\% compared to the other 6 models.
 
\begin{figure}[h]
\centering
    \includegraphics [width=0.4\textwidth]{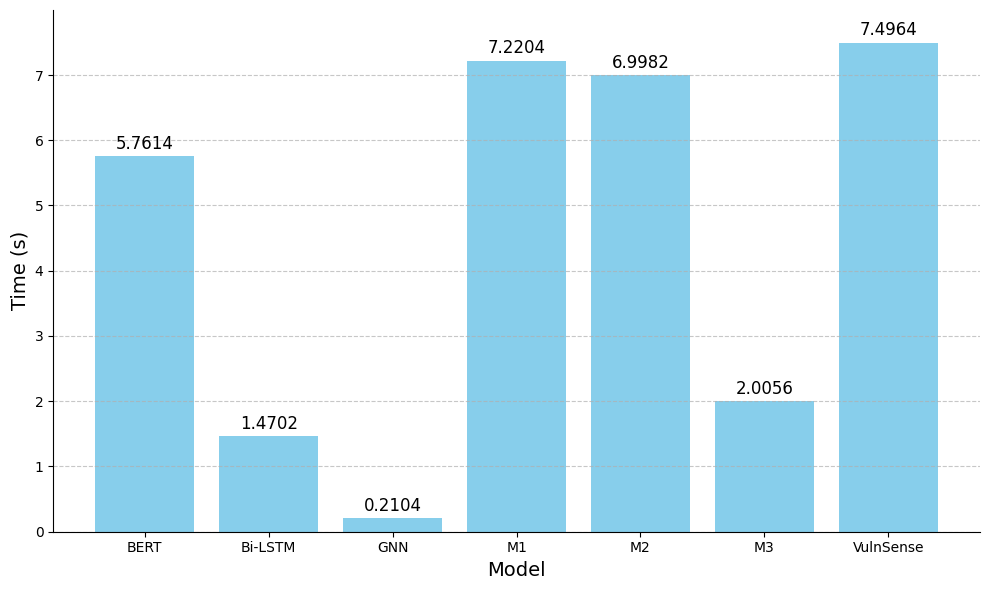}
    \caption{Comparison chart of prediction time on the test set for each model}
    \label{fig:cmpTimePredict}
\end{figure}

In addition, \textbf{Figure \ref{fig:cmpTimePredict}} illustrates the prediction time on the same testing set for each model. It's evident that these multimodal models M1, M2, and VulSense, which are incorporated from BERT, as well as the unimodal BERT model, exhibit extended testing durations, surpassing 5.7 seconds for a set of 354 samples. Meanwhile, the testing durations for the GNN, BiLSTM, and M3 models are remarkably brief, approximately 0.2104, 1.4702, and 2.0056 seconds correspondingly. It is noticeable that the presence of the unimodal models has a direct influence on the prediction time of the multimodal models in which the unimodal models are involved. In the context of the 2 most effective multimodal models, M3 and VulSense, M3 model gave the shortest testing time, about 2.0056 seconds. On the contrary, the VulSense model exhibits the lengthiest prediction time, extending to about 7.4964 seconds, which is roughly four times that of the M3 model. While the M3 model outperforms the VulSense model in terms of training and testing duration, the VulSense model surpasses the M3 model in accuracy. Nevertheless, in the context of detecting vulnerability for smart contracts, increasing accuracy is more important than reducing execution time. Consequently, the VulSense model decidedly outperforms the M3 model.

\section{Conclusion} \label{conclusion}

In conclusion, our study introduces a pioneering approach, VulnSense, which harnesses the potency of multimodal deep learning, incorporating graph neural networks and natural language processing, to effectively detect vulnerabilities within Ethereum smart contracts. By synergistically leveraging the strengths of diverse features and cutting-edge techniques, our framework surpasses the limitations of traditional single-modal methods. The results of comprehensive experiments underscore the superiority of our approach in terms of accuracy and efficiency, outperforming conventional deep learning techniques. This affirms the potential and applicability of our approach in bolstering Ethereum smart contract security. The significance of this research extends beyond its immediate applications. It contributes to the broader discourse on enhancing the integrity of blockchain-based systems. As the adoption of smart contracts continues to grow, the vulnerabilities associated with them pose considerable risks. Our proposed methodology not only addresses these vulnerabilities but also paves the way for future research in the realm of multimodal deep learning and its diversified applications. 

In closing, VulnSense not only marks a significant step towards securing Ethereum smart contracts but also serves as a stepping stone for the development of advanced techniques in blockchain security. As the landscape of cryptocurrencies and blockchain evolves, our research remains poised to contribute to the ongoing quest for enhanced security and reliability in decentralized systems.


%

\section*{Acknowledgment}


This research was supported by The VNUHCM-University of Information Technology's Scientific Research Support Fund.


 \bibliographystyle{elsarticle-num} 
 \bibliography{cas-refs}





\end{document}